# TOY MODEL OF THE NUCLEON – NUCLEON POTENTIAL

G. P. Kamuntavicius, A. Masalaite, S. Mickevicius

*Vytautas Magnus University, Vileikos 8, Kaunas 44404, Lithuania*

We study the simplest quark model, assumed that the sea of gluons and quark – antiquark pairs could be treated as part of a static force, and proceeded to calculate the hadronic states by solving the Schrodinger equation for a static confining interaction. We refer to this model, starting from a system of six interacting constituent quarks and examine, how the picture of two structureless nucleons can change when the effects caused by the substructure of the nucleons are taken into account.

## 1. Introduction

One of the fundamental goals of nuclear theory is to explain the properties of atomic nuclei in terms of the elementary interactions between pointlike nucleons. By construction, nucleon - nucleon (*NN*) potentials must, first of all, reproduce the two - nucleon scattering data and the properties of the deuteron. Recently, progress has been made not only in the phase - shift analysis, but also in the accuracy and consistency of the fits of realistic *NN* potentials to these data. As a result, several charge - dependent *NN* potentials have been constructed, which give a very reasonable fit to the energy - dependent partial - wave analyses of the *NN* scattering and produce very good description of deuteron and these *NN* data below 350 *MeV*. Potentials, like the recent Nijmegen (Nijm I, Reid93, and Nijm II) [1], the Argonne (AV18) [2], and the CD Bonn [3], yield a $\chi^2$/datum of about 1 and may be called phase - shift equivalent. Moreover, the potentials Reid93, Nijm II and AV18 are local potentials. These potentials enjoy great popularity, because they are easy to apply in configuration - space calculations. As such, they are the best candidates for *NN* potentials to use in calculations of nuclei having more than two nucleons.

Unfortunately, good *NN* potentials, defined as above, can not reproduce even the binding energy of the three - nucleon systems. All local realistic *NN* forces underbind the triton by some 0.8 *MeV*. Similarly, the $\alpha$ - particle and the lightest nuclei (with $A \leq 7$), for which more or less accurate solutions of the Schrodinger equation can be performed, are all underbound by these potentials. So far, a few different ways for solving this discrepancy have been presented and investigated. They are (1) the relativistic corrections, (2) a nonlocal *NN* potential, (3) three - nucleon forces and (4) the structure of nucleons taken into account.

Let us briefly summarize the main results of these modifications. Fully relativistic calculations are extremely complicated and consequently have not yet been carried out. The kinematical corrections, yielding a Hamiltonian with the correct transformation properties up to order $(v/c)^2$ produces results that are small and repulsive: approximately 0.3 *MeV* of repulsion in the triton and almost 2 *MeV* in the *α* - particle [4]; see, however [5]. Nonlocal *NN* potentials, such, as the CD Bonn, can improve the result for the binding energy of the triton by some 0.4 *Me*V, but not more [6]. The most impressive results for solving the problem of underbinding are obtained by applying phenomenological three - nucleon forces, adjusted to achieve the correct triton ground - state energy [7]. With this addition $^4$He is properly bound, while the ground states energies of *A* = 5 - 8 and the excitation energies of the low - lying states are again too high [8].

Many studies have been devoted so far to the understanding of the *NN* interaction starting from quarks models. A systematic connection to quantum chromodynamics is established by chiral effective field theory. Up to now the two - nucleon system has been considered in chiral perturbation theory [9,10]. However, due to the formidable mathematical problems, we are still far from a quantitative understanding of the *NN* force from this point of view. Nevertheless the success of the pointlike constituent quark model in barion spectroscopy demonstrates, that successful application of this approach for two nucleons could be the best chance to understand peculiarities of this interaction and to solve the problem, how *NN* potential modifies in nuclei due to presence of surrounding nucleons. Let us refer to the simplest quark model, starting from a system of six constituent quarks and examine now, how the picture of two structureless nucleons can change when the effects caused by the substructure of the interacting nucleons are taken into account.

## 2. Confining wells of the two – nucleons system

The picture of structureless nucleons keeping individuality in bound state of two nucleons (deuteron) and scattering states up to 350 *MeV*, so successful for realistic potentials definition, is in essence not consistent with any known scenario with six interacting quarks involved into play. The models based on one – gluon exchange between quarks can explain only the short – range repulsion of the *NN* potential. The middle and long – range attraction is obtained from the meson exchanges between quarks [11]. The way we can present solution of this problem is nontraditional consideration of problems with confinement in case, when two nucleons approach one another. As it is well - known, the established low energy spectrum of quantum chromodynamics behaves as though hadrons are dominated by their valence quark structure and confinement. Also from the point of view of quarks all nuclei are confined too. The problem is how the idea of quarks, confined in nucleons, can be applied for nuclei. Let the quarks are trapped in nucleons by infinitely deep confining harmonic oscillator (HO) potential, as it is often used in the Standard Model. When nucleons approach each other, the first nucleon confinement potential comes into contact with the corresponding potential of the second nucleon. Let us for the sake of simplicity start consideration with one - dimensional harmonic confining wells for quarks with point of contact $z = 0$. Nucleons are identical, so these confinement potentials are symmetrical. Let bottoms of wells are situated in points $z_0$ and $-z_0$ respectively. The essential and original our suggestion is that than in case when nucleons go into contact overlaping confining wells must vanish. An example of such a potential, corresponding to $z_0 = 1$ is given in Fig. 1.

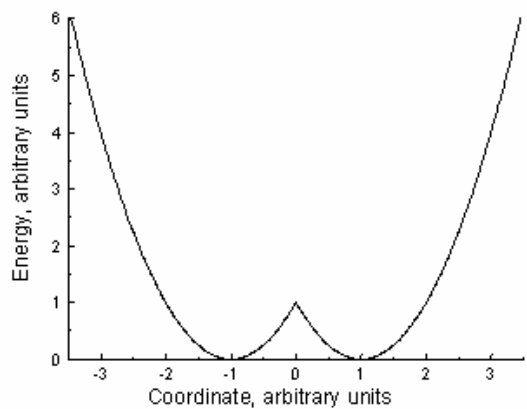

Fig.1. The confinement potential for two nucleons.

The left - hand side of this potential well can be expressed analytically as $V(z) = \dfrac{m\omega^2(z+z_0)^2}{2}$, where $m$ is the quark mass (taken equal to 1/3 of the nucleon mass) and $\omega$ is the HO frequency. In the same way the right side of the potential equals $V(z) = \dfrac{m\omega^2(z-z_0)^2}{2}$. Applying the Heviside function:

$$\Theta(z) = \begin{cases} 0, & z < 0, \\ \frac{1}{2}, & z = 0, \\ 1, & z > 0 \end{cases} \quad (1)$$

one can define analytical expression for the potential from Fig. 1 as

$$V(z) = \frac{m\omega^2 \left[(z-z_0)^2 \Theta(z) + (z+z_0)^2 \Theta(-z)\right]}{2}, \quad (2)$$

or

$$V(z) = \frac{m\omega^2 (|z|-z_0)^2}{2}. \quad (3)$$

The Hamiltonian for quark, moving in this well is given by

$$H(z) = -\frac{\hbar^2}{2m}\frac{d^2}{dz^2} + \frac{m\omega^2(|z|-z_0)^2}{2}. \quad (4)$$

Thus far one considered one - dimensional confinement potential, while quarks in nucleons move in the three – dimensional well. The three - dimensional Schrodinger equation for constituent quark is:

$$\left[-\frac{\hbar^2}{2m}\nabla_{\mathbf{r}} + \frac{1}{2}m\omega^2(\mathbf{r}-\mathbf{r_0})^2\Theta(z) + \frac{1}{2}m\omega^2(\mathbf{r}+\mathbf{r_0})^2\Theta(-z) - E\right]\Psi(\mathbf{r}) = 0. \quad (5)$$

Let us introduce the dimension - free variables and overwrite the Hamiltonian as:

$$\left[-\nabla_{\mathbf{r}} + (\mathbf{r}-\mathbf{r_0})^2\Theta(z) + (\mathbf{r}+\mathbf{r_0})^2\Theta(-z) - E\right]\Psi(\mathbf{r}) = 0. \quad (6)$$

Here dimension – free coordinates are given in terms of oscillator length parameter $b$ and energy is given in coresponding energy quanta $\hbar\omega$. As $\hbar\omega \cdot b^2 = \frac{\hbar^2}{m}$, the constituent quark mass $m$, equal the one third of the nucleon mass $\left(m = \frac{m_{nucleon}}{3}\right)$, defines the relation of both parameters introduced: $\hbar\omega \cdot b^2 = 125\, MeV\, fm^2$.

Taking centers of confining wells for nucleons in points $(0, 0, z_0)$ and $(0, 0, -z_0)$ one can simplify the confining potentials due to relations:

$$(\mathbf{r}-\mathbf{r_0})^2 = x^2 + y^2 + (z-z_0)^2 \quad (7)$$

and

$$(\mathbf{r}+\mathbf{r_0})^2 = x^2 + y^2 + (z+z_0)^2. \quad (8)$$

Obviously, the wells overlap and vanish on (x, y) plane of introduced coordinate system hence one can present Eq. (6) in form:

$$\left[-\frac{d^2}{dx^2}+x^2-\frac{d^2}{dy^2}+y^2-\frac{d^2}{dz^2}+(z-z_0)^2\Theta(z)+(z+z_0)^2\Theta(-z)-E\right]\Psi(x,y,z)=0. \quad (9)$$

So, the variables in the Schrodinger equation can be separated and its enough to consider the one – dimensional potential along $z$ - axis, i.e. along the line connecting center of mass of interacting nucleons.

The eigenvalues and eigenfunctions of corresponding one - dimensional Hamiltonians in separate confining wells are

$$E_n = n+\frac{1}{2}, \quad \Psi_n(z) = \sqrt{\frac{1}{2^n n!\sqrt{\pi}}} H_n(z) e^{-\frac{z^2}{2}}, \quad (10)$$

where $H_n(z)$ are Hermite polynomials of $n$ – th order. So, the eigenfunctions of the left – hand side Hamiltonian, present in Fig. 1 are

$$\Psi_n(z+z_0) = \sqrt{\frac{1}{2^n n!\sqrt{\pi}}} H_n(z+z_0) e^{-\frac{(z+z_0)^2}{2}}, \quad (11)$$

and the corresponding right – hand side eigenfunctions are

$$\Psi_n(z-z_0) = \sqrt{\frac{1}{2^n n!\sqrt{\pi}}} H_n(z-z_0) e^{-\frac{(z-z_0)^2}{2}}. \quad (12)$$

The solutions of the Schrodinger equation in the joint confining well exist, when eigenfunctions (11) and (12) and their derivatives satisfy the continuity conditions at point $z = 0$:

$$\Psi_n(z-z_0)\big|_{z=0} = \Psi_n(z+z_0)\big|_{z=0} \quad (13)$$

and

$$\Psi'_n(z-z_0)\big|_{z=0} = \Psi'_n(z+z_0)\big|_{z=0}. \quad (14)$$

Namely these conditions for wave-functions give the quantization of spectrum in joint confining well as function of parameter $z_0$. The values of oscillator quanta $n$ versus parameter $z_0$, at which the wave - funtions satisfy conditions (13) and (14), are given in Fig. 2 and Fig. 3.

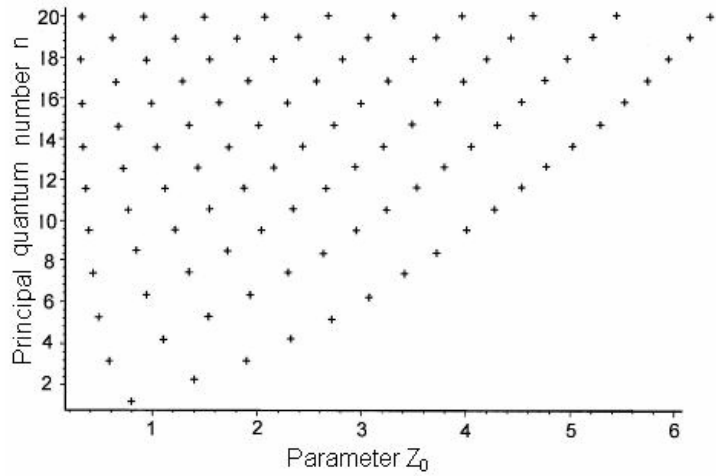

Fig. 2. Values of the oscillator quantum number *n* versus parameter $z_0$ at which the wave-functions $\Psi_n(z_0) = \Psi_n(-z_0) = 0$ and derivatives of these functions satisfy the continuity condition (14).

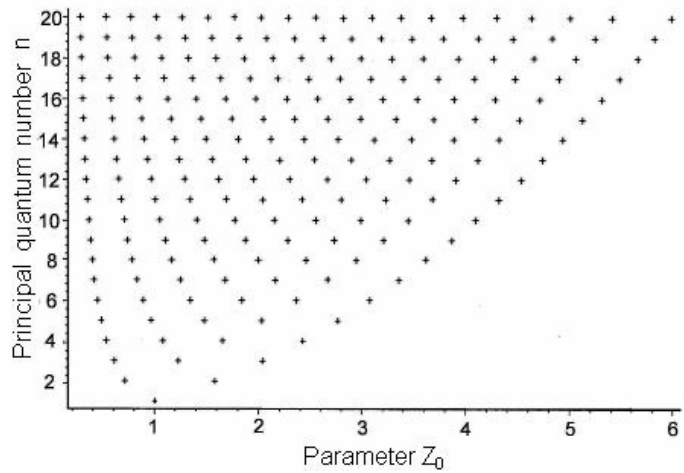

Fig. 3. Values of the oscillator quantum number *n* versus parameter $z_0$ at which the wave-functions satisfy the continuity condition (13) and derivatives of these functions $\Psi'_n(z_0) = \Psi'_n(-z_0) = 0$.

Let us demonstrate some wave-functions, matching these conditions. As mentioned, the principal quantum number *n* gives the number of zeros of corresponding wave-function in own (separate) confining well. Even values of this quantum number corresponds to functions symmetrical in own well, odd values – to functions antisymmetrical in this well. Eigenfunctions of joint well Hamiltonian, symmetrical as well as antisymmetrical, can be composed as of functions, symmetrical in own wells, as well as of antisymmetrical functions. The functions, antisymmetrical in joint well equals zero in center of well, i.e. in point $z = 0$ (Fig. 2), symmetrical functions have local maximum or local minimum (zero derivative) at this point (Fig. 3).

An examples of the wave - functions, both antisymmetrical and symmetrical in joint confining wells are shown correspondingly in Fig. 4 and Fig. 5. The energy of bound state in joint confining well is indicated by horizontal line. This line is serve as zero line for wave – function present.

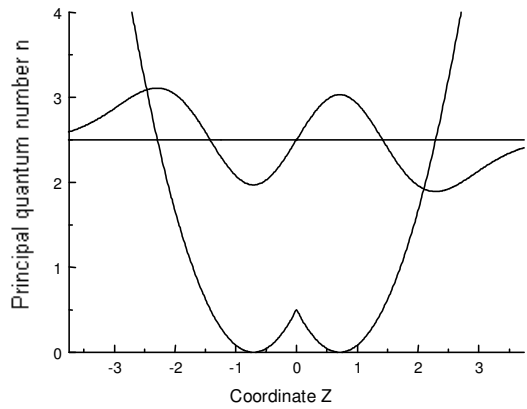

Fig. 4. An example of the wave - function, antisymmetrical in joint confining well.

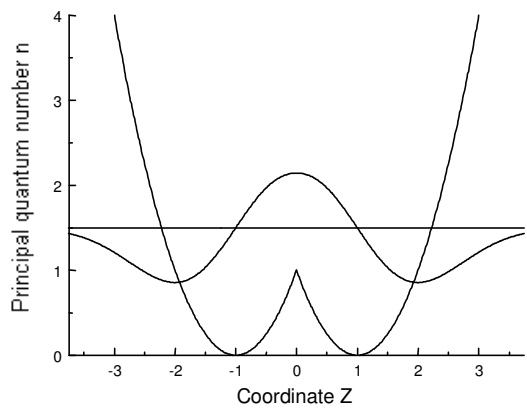

Fig. 5. An example of the wave - function, symmetrical in joint confining well.

As one can study from figures present, at any small, not equal to zero $z_0$ value, the ground state of this joint confining well is situated at very high energy. At growing $z_0$ the ground state in the well moves down. At some value of $z_0$ it reaches minimal value and starts moving up. Finally, at some value of $z_0$ it appears the possibility for quarks from two separated nucleons to occupy the Standard Model states $n = 0$ in different confining wells.

By a way, this function is not an exact eigenfunction for this joint well, but overlap of functions from different wells is negligible, so from quantum mechanical point of view these states are allowed. When $z_0 = 3.19$ two lowest levels in joint well are $n = 9$ (exact level), and $n = 0$ (approximate level), shown in Fig. 6.

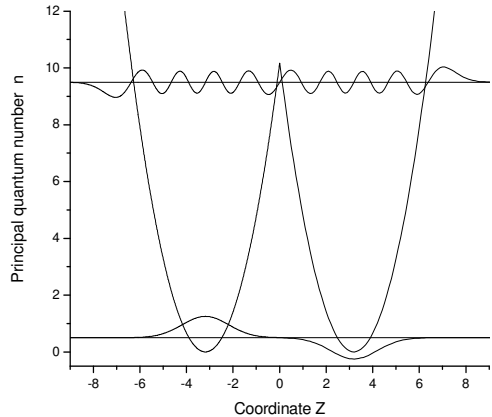

Fig. 6. An example of two lowest levels in joint well. The level $n = 0$ is approximate level and represents quarks from two separated nucleons. The level $n = 9$ is exact level and represents excited quark state.

Let us estimate the parameters of the model. The relation between $b$ and the nucleon radius is $\sqrt{\frac{3}{2}}b = 0,8\,fm$. This gives $b = 0,65\,fm$. This value is close to value for this parameter of nonrelativistic quark model, equal $0,5\,fm$. The distance between centers of nucleons, corresponding to the attraction, equals $2z_0 b = 1,4\,fm$ (Fig. 2). This value is very close to distance, at which is situated a bottom of the potential well of local realistic potentials. The corresponding value for HO energy quantum is $\hbar\omega = \dfrac{125\,MeVfm^2}{(0,65\,fm)^2} = 292,96\,MeV$. It approximately equals two pions energy, exchange of which is responsible for interaction between nucleons in the area of bottom of the potential well.

### 3. Conclusions

This oversimplified model, operating with confining wells at small distances between centers of mass of nucleons, excities them and produces high excitations in six quark system hence provides the short - range repulsion necessary to reproduce the experimental data without any need of constituent quark Pauli principle and one - gluon exchange taken into account. To some extent it corresponds to "quark soap" scenario. No one individual nucleon is obtainable at these distances. At large values of $z_0$ the situation changes and it appears the possibility of some individualization of clusters (nucleons) in six quark system. So, the introduced modification of confinement potential in six quark system is well consistent with characteristic features of realistic potentials of *NN* interaction (core, attraction region and asymptotic part) and gives the possibility for modification of *NN* potential, when interacting nucleons are surrounded by additional (spectator) nucleons, i. e. when they are in nuclei.